# PROBLEMS TO BE CLARIFIED BY MEANS OF RADIOACTIVE ION BEAMS PROVIDED BY THE ACCULINNA-2 SEPARATOR

G.M. Ter-Akopian[1,2], A.A. Bezbakh[1,3], V. Chudoba[1,8], A.S. Fomichev[1],
M.S. Golovkov[1,2], A.V. Gorshkov[1,3], L.V. Grigorenko[1,4,5], G. Kaminski[1,6],
A.G. Knyazev[1], S.A. Krupko[1,3], M. Mentel[1,7], E.Yu. Nikolskii[1,4], Yu.L. Parfenova[1],
P. Pluczinski[1,7], S.A. Rymzhanova[1,3], S.I. Sidorchuk[1], R.S. Slepnev[1], S.V. Stepantsov[1],
R. Wolski[1,6], B. Zalewski[1,9]

[1] *G.N. Flerov Laboratory of Nuclear Reactions, JINR, 141980 Dubna, Russia*
[2] *National University 'Dubna', 141980 Dubna, Russia*
[3] *SSC RF ITEPh of NRC "Kurchatov Instutute", 117218 Moscow, Russia*
[4] *NRC "Kurchatov Institute", Kurchatov sq 1, 123182 Moscow, Russia*
[5] *NR Nuclear University "MEPhI", Kashirskoe shosse 31, 115409, Moscow, Russia*
[6] *Henryk Niewodniczanski Institute of Nuclear Physics, PAS, 31342 Cracow, Poland*
[7] *AGH University of Science and Technology, Faculty of Physics and Applied Computer Science, al. Mickiewicha 30, 30059 Cracow, Poland*
[8] *Institute of Physics, Silesian University, 74 601 Opava, Czech Republic*
[9] *Heavy Ion Laboratory, Univerity of Warsaw, 02-093 Warsaw, Poland*

Fragment separator Acculinna 2 has been built in the G.N. Flerov Laboratory of Nuclear reaction (JINR, Dubna). It is timely now to choose meaningful and challenging objectives for experiments dedicated to the study of light drip-line nuclei. Considerable interest makes the search for the minor 2p-decay branch of the first excited state of $^{17}$Ne. The knowledge on the $\Gamma_{2p}/\Gamma_\gamma$ width ratio for this excited state is of considerable interest because the reverse process of simultaneous two-proton capture could be a bypass for the $^{15}$O waiting point occurring in the CNO cycle of nucleosynthesis. Accumulation of high-statistics data for the $^{10}$He excitation spectrum populated in the $^{2}$H($^{8}$He,p)$^{9}$He and $^{3}$H($^{8}$He,p)$^{10}$He reactions, as well as the study of cross-check reactions made with the $^{11}$Li and $^{14}$Be beams, will make an effective way to clarify the succession of $^{10}$He excited states. A hot topic beyond the neutron drip line makes the observation of results capable to elucidate the low-energy resonance states anticipated for the 4n decay of $^{7}$H. The RIB beams provided by Acculinna-2 will allow one to perform experiments where luminosity coming to a level of more than $2\times10^{26}$ cm$^{-2}$ s$^{-1}$ will be achievable in experiments aimed study of the $^{2}$H($^{8}$He,$^{3}$He)$^{7}$H and $^{2}$H($^{11}$Li,$^{6}$Li)$^{7}$H reactions.

## 1. Introduction.

Radioactive ion beams (RIBs) supplied by the Acculinna [1] separator proved their worth as reliable tool in the study of light drip-line nuclei carried out in Dubna during the last 20 years [2]. Though Fig. 1 shows a relatively small part of the Nuclide Map it makes a topic of challenging work in the contemporary nuclear physics. For nuclei laying both on the neutron and proton drip lines stability limits are achieved here and are the subject of study just in this locality where new aspects of nuclear dynamics become evident (see review [3] and references therein).

In the vicinity of drip lines clustering becomes a common phenomenon in these exotic nuclei; here the notion of nuclear-density saturation works no more. Therefore one should make difference between the motion dynamics of clusters and nucleons bound in these clusters. Nuclei with far extended neutron haloes are obtained close to the neutron drip line. Some notable one-neutron ($^{11}$Be) and two-neutron ($^6$He, $^8$He, $^{11}$Li, $^{14}$Be) halo nuclei are indicated in Fig. 1. Inherent to the halo nuclei is the emergence of the so-called soft excitation modes.

Considering the neutron drip-line nuclei we notice that versatile and reliable data should be obtained elucidating the spectra of $^{10}$He and its neighbors. The breakdown of the N = 8 magic number obtained for the $^{10}$He nucleus makes urgent this study. Another challenge makes the possible existence of novel types of radioactivity, such as the two-neutron and four-neutron radioactive decay which perhaps could be discovered for some nuclei indicated in Fig. 1. In the search for the 2n radioactivity the points of interest make $^{16}$Be and $^{26}$O. The $^7$H and $^{28}$O nuclei are candidate for 4n radioactivity.

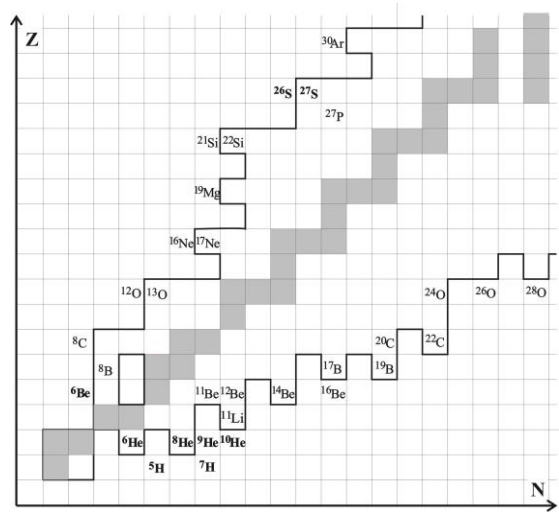

Fig. 1. Light nuclei in the neighborhood of the neutron/proton drip lines. Gray squares outline the stability valley.

Pairing effect sharply comes out at the stability borders resulting in specific effects, i.e. to the appearance of borromean nuclei, the two-proton radioactivity and the true three body (democratic) decay. The respective proton-excess nuclei are indicated in Fig. 1.

## 2. Distinctive features of experiments at the separator Acculinna-2.

Acculinna-2 is designed and built [4,5] to be a more effective substitute for the Acculinna separator. Preparation for the Acculinna-2 start-up is presented in the talk given by S.A. Krupko at this symposium. Possibilities offered by the new separator in the studies of light exotic nuclei will be discussed here.

The primary heavy-ion beam of the U400M cyclotron is focused on the production target located at the beginning of the main separator section. This is an achromatic ion-optical structure having one dispersive focal plane where a wedge-shaped beryllium degrader is placed. The wanted RIB nuclei, cleaned from the primary beam ions and from the majority of other reaction products, are focused on final achromatic plane situated outside the cyclotron hall in a room free of radiation created by the primary beam. Guided by the quadrupole magnets of the separator third section, the RIB nuclei run to the physics target. A long 14-meter flight base is offered here. It is proper for the time-of-flight measurement giving the energy of individual RIB nuclei ascertained with one-percent accuracy. Also, the magnetic structure of this section is optimized for the installation of an RF filter which will be manufactured later and installed at the proper location. Position sensitive chambers (MWPC or PPAC) installed at the end part of this section will measure the inclination angles of individual nuclei hitting the target with accuracy 3 – 4 mrad.

Table 1. Characteristics expected for RIBs obtained from Acculinna-2.

| Primary beam | | Radioactive Ion Beam (RIB) | | | |
|---|---|---|---|---|---|
| Ion | Energy (MeV)/u | Ion | Energy (MeV/u) | Intensity [1] ($s^{-1}$) | Purity (%) |
| $^{11}$B | 32 | $^{8}$He | 26 | $3\times10^5$ | 90 |
|  | 55[2] | $^{8}$He | 42 | $1.5\times10^6$ | 100 |
| $^{15}$N | 49 | $^{8}$He | 37 | $2\times10^5$ | 99 |
|  |  | $^{11}$Li | 37 | $3\times10^4$ | 95 |
| $^{11}$B | 32 | $^{10}$Be | 26 | $1\times10^8$ | 90 |
| $^{15}$N | 49 | $^{12}$Be | 38.5 | $2\times10^6$ | 70 |
| $^{18}$O | 48[2] | $^{14}$Be | 35 | $2\times10^4$ | 50 |
| $^{18}$O | 33 | $^{14}$Be | 26 | $6\times10^3$ | 50 |
| $^{22}$Ne | 44 | $^{17}$C | 33 | $3\times10^5$ | 40 |
|  |  | $^{18}$C | 35 | $4\times10^4$ | 30 |
| $^{36}$S | 64[2] | $^{24}$O | 40 | $2\times10^2$ | 3 |
|  |  |  |  |  | 20[3] |
|  |  | $^{30}$Ne | 45 | $4\times10^1$ | 3 |
| $^{40}$Ar | 67[2] | $^{28}$Ne | 44 | $1\times10^3$ | 10 |
|  |  | $^{27}$F | 45 | $2\times10^1$ | 2 |
| $^{10}$B | 39 | $^{7}$Be | 26 | $8\times10^7$ | 90 |
| $^{20}$Ne | 53 | $^{18}$Ne | 34 | $2\times10^7$ | 40 |
|  |  | $^{17}$Ne | 35 | $1\times10^6$ | 10 |
|  |  | $^{15}$O | 35 | $3\times10^7$ | 60 |
| $^{32}$S | 52 | $^{28}$S | 31 | $2\times10^4$ | 5[3] |
|  |  | $^{24}$Si | 32 | $2\times10^4$ | 5[3] |

[1] RIB intensities are given with reference to a 1 pμA primary beam.
[2] Will be available after the modernization of U400M cyclotron planned for 2019.
[3] Will be obtained with the RF filter installed at the third section of Acculinna-2.

Increased apertures of magnets employed in the new separator offer RIB intensities increased by 30 times as compared to those obtained with Acculinna. Better momentum resolution granted for the separated RIBs suggests their improved purity. Characteristics expected for some RIBs which will be obtained from the Acculinna-2 separator are presented in Table 1. The RIB energies, given there, represent some typical choice available for the set of primary beams accelerated at the U400M cyclotron. The RIB intensities and their purity were calculated using the LISE++ code [6,7].

Among the other, in many respects much more large-scale RIB separators [8-11] Acculinna-2 is determined to give RIBs of intermediate energy especially suitable to carry out experiments aimed at the study of exotic, border-line nuclei produced in transfer reactions. Reactions used here are distinguished for their large, negative Q values. Therefore an upgrade, increasing the primary beam energies is planned for the U400M cyclotron. Favorable for the study of exotic nuclear spectra are the transfer reactions induced when the RIBs are bombarding isotopes of hydrogen ($^1$H, $^2$H, $^3$H) and helium ($^3$He, $^4$He). The reverse kinematics, convenient for the registration of transfer-reaction recoils, makes the missing-mass measurements more easy and precise. But more important are the novel approaches inaccessible in the case of knock-out reactions being in use widely in experiments employing relativistic RIBs.

One well proved novel approach is the correlation analysis of reaction products emitted in the transfer reactions. This provides unequivocal spin-parity ascertainment made for the wide resonant states typically populated when the subject of study is related to exotic drip-line nuclei. The full-scale power of this method was confirmed in the study made on the low-energy $^5$H spectrum [12]. Another invention is the so-called combined-mass method [3]. The main points of this method will be discussed below (see also the paper presented to these proceedings by P.G. Sharov et al.).

3.   **Principal experiments with the Acculinna-2 RIBs.**

The start-up of Acculinna-2 is planned for 2017. It supposes RIB production and delivery to the focal plan where the targets bombarded by RIBs and detector arrays will be installed. A probable choice for the first, demonstration experiment is the $^2$H($^{10}$Be,$^6$Li)$^6$He reaction study. Taking the $^{10}$Be projectile is realistic as a quite good intensity of this RIB requires a low, ~1 pnA current of the primary $^{11}$B beam (see Table 1). Data accumulated on the cross section of α-cluster transfer from $^{10}$Be, and level population in $^6$He, will be reasonable as the first phase in the study of this class reaction in the vicinity of the neutron drip line. Particularly appealing will be the $^2$H($^{14}$Be,$^6$Li)$^{10}$He reaction leading to the key-interest helium isotope $^{10}$He. After these tests more sophisticated research will be timely. Some priority tasks are considered below.

*3.1 Search for the 2p decay branch of the first excited state in $^{17}$Ne.*

The energy of the 1288-keV, $J^\pi = 3/2^-$ first excited state of $^{17}$Ne exceeds the threshold of 2p emission by 344 keV, whereas the proton emission from this state is not allowed. Therefore the true two-proton decay (see review paper [13]) might be possible if the 2p-decay branch of this $^{17}$Ne state could compete with γ decay. The experimental measurement of the $\Gamma_{2p}/\Gamma_\gamma$ width ratio made for this excited state is of considerable interest because the reverse process of simultaneous two-proton capture by $^{15}$O could be a bypass for this waiting point occurring in the CNO cycle of nucleosynthesis [14].

Searches made before for the 2p decay mode of this $^{17}$Ne state gave an experimental limit $\Gamma_{2p}/\Gamma_\gamma \leq 7.7\times 10^{-3}$ [15], while the theory work [16] predicts a value of $\Gamma_{2p}/\Gamma_\gamma \approx 2\times 10^{-6}$. A careful analysis of data presented at EXON 2014 [17] shows that in our work intended to test the noticed above combined-mass method the experimental limit obtained for the first 3/2$^-$ state of $^{17}$Ne is lowered to a level of $\Gamma_{2p}/\Gamma_\gamma \leq 5\times 10^{-4}$.

Approach which we call combined-mass method is clarified well on the example of the $^{1}$H($^{18}$Ne,d)$^{17}$Ne reaction. The emission angle and energy measured for the recoil deuteron make the objective when the excitation spectrum of $^{17}$Ne is measured by the missing mass method. The yields of different resonance states are defined by this method irrespective of their decay mode. The detection of the recoil deuteron made in coincidence with protons offers a typical way to the yield determination made for the decay branches of the $^{17}$Ne excited states associated with the proton emission. Certainly, of particular interest here is the search for the anticipated very weak two-proton decay branch of its first exited state (E*=1288 keV, J$^\pi$=3/2$^-$). Having the emission angle of the recoil deuteron measured even with a modest (about one degree) accuracy one specifies with a tremendous accuracy ($\Delta E/E \approx 6\times 10^{-4}$ and 0.1 degree, respectively) the energy and escape direction of $^{17}$Ne in lab system. This is typical in general for the study of transfer reactions made in inverse kinematics with heavy projectiles bombarding light target nuclei. Resolution attainable by means of combined-mass method depends on the target thickness which can be set to be quite large due to the small specific energy losses of the protons which due to detection. This favors the revelation of such small 2p decay branch as it is anticipated for the 1288-keV excited state of $^{17}$Ne.

Principal setups used in the test experiment [17] were just those which were on our disposal. But despite the lack of perfection inherent to its implementation this work demonstrated that the combined-mass method provides condition for the further progress in attempts aimed at revealing transitions characterized by the branching ratio being as low as $\Gamma_{2p}/\Gamma_\gamma \approx (1 - 2)\times 10^{-6}$. For the 2p branch of the first excited state of $^{17}$Ne this little, but still measurable, $\Gamma_{2p}/\Gamma_\gamma$ implies that the reaction channel having cross section $d\sigma/d\Omega \approx 1$ nb/sr becomes the subject of study.

### 3.2 Breakdown of shell closure in $^{10}$He.

In former times it could be natural to take that $^{10}$He is the next, after $^{4}$He, double magic nucleus. This opinion stimulated numerous experiments carried out in attempts to obtain this helium isotope. The first paper reporting the $^{10}$He discovery was published in 1994 [18]. Bombarding targets of deuterated polyethylene and pure graphite with 61 MeV/u $^{11}$Li beam the authors studied the $^{8}$He +n +n coincidence events. A wide ($\Gamma \sim 1.2$ MeV) peak centered at about 1.2 MeV above the 2n decay threshold was obtained in the invariant mass spectrum built out of these events. This peak was explained by the authors as the $^{10}$He ground-state resonance. Later on, wide $^{8}$He +n +n peaks with energy 1 – 1.5 MeV obtained in fragmentation reactions of $^{11}$Li [19, 20] and $^{14}$Be [21] were ascribed to the $^{10}$He ground-state resonance. But it turned out that another interpretation related to the reaction mechanism is possible for these types of $^{10}$He spectrum population [22, 23].

In our experiments [24, 25] the $^{10}$He nucleus was produced in the $^{3}$H($^{8}$He,p)$^{10}$He reaction. The obtained missing-mass spectrum of $^{10}$He is shown in Fig. 2. Alongside with the recoil proton the $^{8}$He nucleus emitted in coincidence was detected. This ensured knowledge about the $^{8}$He emission angle ($\theta_{8He}$) and about the part ($\varepsilon = E_{nn}/E_T$) of the $^{10}$He center-of-mass total decay energy ($E_T$) going to the relative motion in the 2n subsystem ($\theta_{8He}$ is measured in respect to the momentum-transfer vector obtained from the proton emission angle).

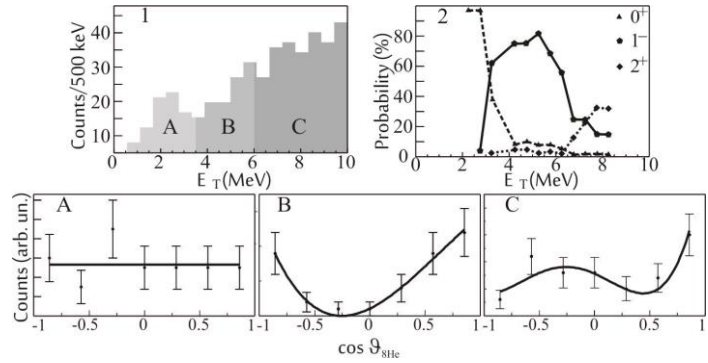

Figure 2. Diagrams demonstrating conclusions following from the study made on the 2n transfer reaction $^{3}$H($^{8}$He,p)$^{10}$He. 1) Missing mass spectrum obtained for the $^{10}$He nucleus; 2) Squared amplitudes of coherent s-, p-, d-wave contributions deduced from the angular distributions of $^{8}$He appearing at the $^{10}$He decay. The lower three panels show these angular distributions averaged over the energy ranges marked A, B, and C in panel 1.

Ranges marked A, B, and C in the $^{10}$He spectrum are characterized by strongly pronounced angular and energy correlations. Analysis done for these correlations made us sure that the wide ($\Gamma > 1$ MeV) $0^+$ ground-state resonance of $^{10}$He is located at $E_T \approx 2.2$ MeV. This state appears being practically pure in the measured $^{10}$He spectrum. In a range of $4.5 < E_T < 6.0$ MeV mainly the first excited $1^-$ state of $^{10}$He contributes in the measured spectrum, and at $E_T > 6.0$ MeV the yield of the next $2^+$ state grows. The appearance of the $1^-$ intruder state in the $^{10}$He spectrum is evidence that the closed-shell structure breaks down in the case of this "doubly-magic" nucleus.

The significance of getting more knowledge about the $^{10}$He structure, including the deeper insight into its excitation spectrum, makes the relevant experiments to be in the priority list of works planned for the RIBs provided by Acculinns-2. This will be the study of the $^{3}$H($^{8}$He,p)$^{10}$He reaction aimed to acquire data exceeding by two orders of magnitude the statistics presented in Fig. 2. Then, the value data obtained for $^{10}$He will get that obtained in the study of the spectrum of the $^5$H nucleus [26]. Important feature of this study was that a high-statistics, complete-kinematic data set was obtained for the $^5$H spectrum populated in the $^3$H($^3$H,p)$^5$H reaction. The same should be done for the $^{10}$He spectrum populated in the $^3$H($^8$He,p)$^{10}$He reaction.

### 3.3   The $^7$H nuclear system and the 4n-decay problem.

Being obtained reliably $^7$H will be a unique nucleus for its maximum neutron excess which one can imagine. Possibility that a narrow low-lying resonance of $^7$H can be observed in experiments was indicated far ago by Ya.B. Zel'dovich [27]. Now, when one knows about the quite well bound nucleus $^8$He, having cluster structure with α core enclosed in 4n halo (skin), an assumption that the $^7$H system, differing from $^8$He by putting triton at the core position, might be not far of stability seems not much unrealistic.

Attempt [28] to observe the 4n radioactive decay of $^7$H set an upper limit of 1 ns for its lifetime. Searches for the $^7$H ground-state resonance populated in the reactions $^1$H($^8$He,2p)$^7$H [29], $^{11}$Be($\pi^-$,p$^3$He)$^7$H [30], and $^2$H($^8$He,$^3$He)$^7$H [31, 32], $^{12}$C($^8$He,$^{13}$N) [33] were undertaken. The most consistent and sensitive experiment [32] was carried out at the $^8$He beam energy 42 MeV/u. Despite the poor, 1.5-MeV resolution being attainable in this work, a peculiarity observed at ~2 MeV in the measured missing-mass spectrum could be understood as indication for some low-energy state (or states) of $^7$H. Cross section ~30 μb/sr was estimated for the $^7$H resonance populated in the $^2$H($^8$He,$^3$He)$^7$H reaction.

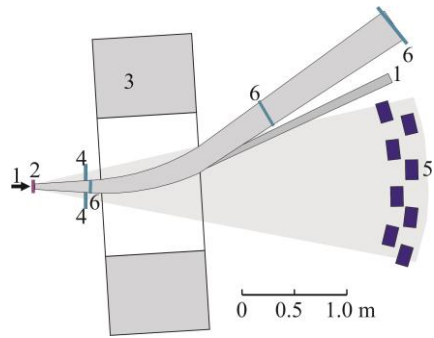

Fig. 3. Layout of setup to be used for the study of a searched low-lying resonance state in $^7$H which could be populated in the $^2$H($^8$He,$^3$He)$^7$H reaction. 1 – $^8$He beam, 2 – deuterium target, 3 – zero-angle dipole magnet, 4 – telescopes detecting $^3$He recoils, 5 – array of stilbene crystals for neutron detection, 6 – array of hodoscopes for coordinate detection of tritons emerging from the $^7$H decay.

Meanwhile, the authors of theory work [34] showed that in the case of simultaneous emission of two or four neutrons (so-called "true" 2- and 4-neutron decays), extremely long (radioactivity scale) lifetimes are possible even for a fairly large decay energies. Search for new phenomenon – multineutron radioactivity – becomes urgent. The arrangement of experiments, aimed for the search for the predicted 4n emitters, calls for serious steps: new methods are needed alongside with a significant improvement in resolution and expansion meant for the achievement higher statistics.

Among the candidates for the 4n radioactive decay ($^7$H, $^{18}$Be, $^{28}$O]) the $^7$H system is within realistic reach for the new ACCULINNA-2 facility and we can join the race for this discovery. The two reactions $^2$H($^8$He,$^3$He)$^7$H and $^2$H($^{11}$Li,$^6$Li)$^7$H are suitable for the study made with the beams of $^8$He and $^{11}$Li obtainable from ACCULINNA-2 (see in Table 1). The energies of these beams, 37 – 40 MeV/u, are optimal for reaching the maximum cross sections available for reactions notable for their rather large negative Q-values. The beam intensities are enough to keep luminosity at a level of ~$2\times10^{26}$ cm$^{-1}$s$^{-1}$ keeping the deuterium targets thin enough for getting at a 500-keV resolution in the measured missing-mass spectra.

The $^7$H problem will be a priority study carried out just after the completion of tests of the new complex. Layout presented in Fig. 3 allows one to catch the concept of these experiments. Besides the detection of recoils emitted in the reactions chosen for the study ($^3$He or $^6$Li recoils assumed to be recorded in the reactions $^2$H($^8$He,$^3$He)$^7$H or $^2$H($^{11}$Li,$^6$Li)$^7$H, respectively) the decay products of $^7$H – the triton and neutrons – will be detected by the setup shown in Fig. 3. The zero-angle dipole magnet, received now from the manufacturer (the Sigma Phi Company [35]), is incorporated into the setup. Hodoscopes placed in three planes along the charged particle trajectories will give information defining the momentum vectors of tritons detected in coincidence with the recoils. The wide aperture of the magnet allows the tritons and neutrons, emitted from the deuterium target, to pass further with minimal losses to the respective detector arrays. The array of scintillation counters shown in Fig. 3 will be equipped with enough number of crystals to bring up the detection efficiency of neutrons to a level no less than 15%.

We anticipate that within 10-day experiments dedicated to the each of these reaction about 300 events will be recorded in the case that the cross section makes 30 μb/sr for the $^7$H system formed with the missing-mass energy $E_{mm}$ < 2 MeV. Each of these events will carry data on the recoil (r), the triton and one or two neutrons detected in coincidence ((r-t-n) or (r-t-n-n) events). The existence of recoil data will give precise knowledge about the flight direction and energy of the formed $^7$H system. Together with the measured time of flight and detection position this makes possible to precisely determine the energy $E_{cm}$ carried out by the individual neutrons running away from the $^7$H center of mass. Resolution attainable in the measurements made for such a $E_{cm}$ spectrum will depend on the region of missing mass energy where this spectrum is observed. Being taken for a range of $E_{mm}$=1–2 MeV the energy $E_{cm}$ carried out by individual neutrons will be defined with errors making about 200 keV. At $E_{mm}$ < 500 keV this error will make 20 – 40 keV. The neutron spectra derived from the (r-t-n) and (r-t-n-n) events will be the subject of theory analysis aimed for the derivation of key parameters of resonance state(s) which will be hopefully obtained in the experiments.

**Acknowledjments.** Work was partly supported by the grant RFBR 14-02-00090.